\documentclass[10pt,letterpaper]{article}
\usepackage{opex3}
\usepackage{color}

\graphicspath{{figures/}}

\begin{document}

\title{Deep learning enabled real time speckle recognition and hyperspectral imaging using a multimode fiber array}

\author{Ulas K\"ur\"um$^1$, Peter R. Wiecha$^1$, Rebecca French$^1$ and Otto L. Muskens$^{1,*}$}

\address{$^1$Physics and Astronomy, Faculty of Engineering and Physical Sciences, University of Southampton, Southampton, UK}

\email{$^*$O.Muskens@soton.ac.uk}

\homepage{https://inanophotonics.southampton.ac.uk/} %% author's URL, if desired

\begin{abstract}
We demonstrate the use of deep learning for fast spectral deconstruction of speckle patterns. The artificial neural network can be effectively trained using numerically constructed multispectral datasets taken from a measured spectral transmission matrix. Optimized neural networks trained on these datasets achieve reliable reconstruction of both discrete and continuous spectra from a monochromatic camera image. Deep learning is compared to analytical inversion methods as well as to a compressive sensing algorithm and shows favourable characteristics both in the oversampling and in the sparse undersampling (compressive) regimes. The deep learning approach offers significant advantages in robustness to drift or noise and in reconstruction speed. In a proof-of-principle demonstrator we achieve real time recovery of hyperspectral information using a multi-core, multi-mode fiber array as a random scattering medium.
\end{abstract}

 % REPLACE WITH CORRECT OCIS CODES FOR YOUR ARTICLE, MINIMUM OF TWO; Avoid using the OCIS codes for ``General'' or ``General science'' whenever possible.
%For a complete list of OCIS codes, visit: http://www.opticsinfobase.org/submit/ocis/

%%%%%%%%%%%%%%%%%%%%%%% References %%%%%%%%%%%%%%%%%%%%%%%%%

%%%%%%%%%%%%%%%%%%%%%%%%%%  body  %%%%%%%%%%%%%%%%%%%%%%%%%%
\section{Introduction}
Motivated by the need for imaging in complex environments and through opaque media, new techniques for characterizing and controlling multiple scattering are currently seeing a tremendous development \cite{rotter_light_2017}. This new toolbox opens up directions for controlling light in random media and exploiting it for applications such as imaging and optical information processing. At the same time, improved understanding allows us to retrieve more information from seemingly random scattering fields to see around corners and through opaque media \cite{katz_looking_2012, bertolotti_non-invasive_2012}. Control over light scattering has led to exciting new applications such as programmable multiport optical elements for classical and quantum states \cite{defienne_nonclassical_2014, huisman_programmable_2015, strudley_ultrafast_2014, park_scattering_2016}, quantum secure keys \cite{goorden_quantum-secure_2014} and compressive sampling imaging systems \cite{liutkus_imaging_2014}.

The multispectral characteristics of speckle fields have been used successfully in a range of studies to realize speckle spectrometers \cite{cao_perspective_2017, redding_using_2012, mazilu_random_2014, chakrabarti_speckle-based_2015, liew_broadband_2016, valley_multimode_2016}. By exploiting wavelength-dependent speckle patterns from a multimode fiber, a spectral resolution of picometers in the near-infrared and nanometers in the visible region has been demonstrated \cite{redding_high-resolution_2014, wan_high-resolution_2015}. Next to single-channel spectrometers, multiplexing of spectrally resolved speckle fields into hyperspectral imaging systems is of great interest. Compared to traditional approaches such as Integral Field Spectrometers \cite{hagen_review_2013, dwight_lenslet_2017}, complex media can offer new opportunities for combining broadband transmission with high spectral resolution \cite{sahoo_single-shot_2017, wang_computational_2018, french_speckle-based_2017}. In the spatial domain, multimode fibers as well as multi-core fiber bundles are a topic of study for a variety of imaging applications such as remote sensing and endoscopy \cite{choi_scanner-free_2012, ploschner_seeing_2015, morales-delgado_two-photon_2015, porat_widefield_2016}.

Recently, a multicore multimode fiber bundle has been used as a frequency characterization element in a high-throughput imaging spectrometer for snapshot spatial and spectral measurements with sub nanometer spectral resolution \cite{french_snapshot_2018}. A compressive sensing (CS) algorithm was successfully employed to retrieve spectral information. Convex regularization techniques such as CS provide a suitable solution for a number of problems in computational imaging. However, in addition to their high computational cost, the applicability depends on a sparsity assumption and indeed performance is reduced for dense data. Faster data processing and requirement of robustness with respect to noise and drift could benefit from an entirely different approach based on artificial neural networks \cite{villmann_processing_2013, zhao_spectralspatial_2016, wang_salient_2016, zhong_spectralspatial_2018, aptoula_deep_2016, li_transferred_2017, yao_intelligent_2018}. Computational methods using neural networks that are trainable for specific problems have recently been shown to be highly efficient and fast \cite{nielsen_neural_2015, goodfellow_deep_2016, wiecha_pushing_2019}. Recently, this approach has been used in various applications utilizing speckle patterns such as image reconstruction, object classification and recognition \cite{yunzhe_deep_2018, satat_object_2017, valent_scatterer_2018, horisaki_learning-based_2016, rahmani_multimode_2018, borhani_learning_2018, moran_deep_2018}.

Here, we demonstrate the successful application of Deep Learning (DL) neural networks to the retrieval of spectral information from speckle images. Using a multi-mode, multicore fiber array as a multiplexed speckle spectrometer, we achieve real-time spectral imaging over several thousands of individual fiber cores. Besides being orders of magnitude faster than other, CS-based techniques, we investigate the robustness of DL to noise as well as to image shifts that could originate from thermal expansion or vibrations in the imaging system. We show the adaptability of DL in such conditions with good performance achieved by appropriate training. Results for DL are compared with CS and with analytical regularized inversion approaches. We find that DL performs well both in the compressive and oversampling regimes, combining a good balance in characteristics with fast reconstruction speeds and massive parallelized performance over many fiber cores.

\section{Method}

\begin{figure}[th]
	\centering\includegraphics[width=10cm]{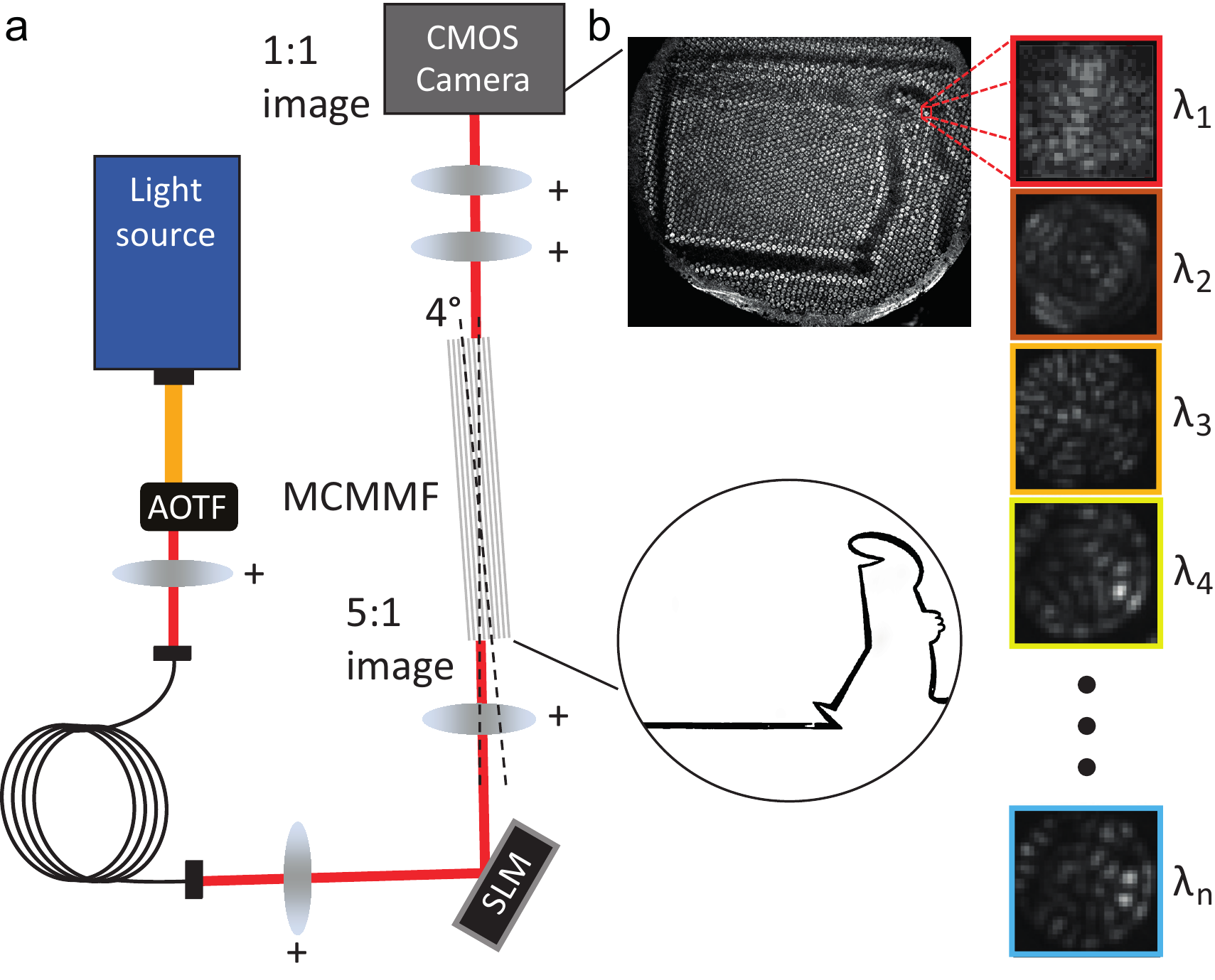}
	\caption{%
		a) Scheme of the experimental setup including broadband supercontinuum laser source, acousto-optical tunable filter (AOTF),  Spatial Light Modulator (SLM) used for image generation and the multi-core, multimode fiber (MCMMF). 
		b) Original projected image and detected camera image of  the exit interface of the fiber bundle at a single selected wavelength with typical speckle patterns of a selected fiber core for different input wavelengths (λ1- λn). La Linea, with permission, copyright CAVA/QUIPOS.}
	\label{fig:fig1}
\end{figure}

We use a multi-core, multimode fiber (MCMMF) as the complex scattering medium (Edmund Optics, Fiber optic image conduit). Mode mixing in each individual fiber results in a characteristic speckle pattern with a wavelength dependence determined by the fiber length and the angle of incidence of the incident light. For a direct comparison with previous results we used in our first calibration experiments the setup as described in \cite{french_snapshot_2018}. For the projection of animations we developed a new setup as shown in Fig.~\ref{fig:fig1}(a) based on an acousto-optic tunable filter (AOTF) and spatial light modulator (SLM).  
In short, a supercontinuum light source (Fianium SC400) was spectrally filtered using an AOTF with a resolution of 5 nm. The filtered light was projected onto a single-mode fiber to ensure stability of the illumination beam with wavelength and to eliminate any other forms of spectral drift in the setup. The fiber output was reflected off the liquid crystal spatial light modulator (SLM, Holoeye Pluto) and was projected onto the MCMMF at an incident angle of 4\(^{\circ}\) with an image demagnification of 5:1. The MCMMF consisted of 3012 fibers with individual core diameters of 50 \textmu m. Fibers of different lengths can be used for different applications depending on the bandwidth required \cite{french_snapshot_2018}. After transmission through the fiber array, the output facet of the MCMMF was imaged onto the focal plane array of a 12-bit, 5 MPixel monochrome CMOS camera with a pixel size of 2.2 \textmu m \(\times\) 2.2 \textmu m (AVT Guppy) using a 1:1 imaging system. Collected images were transferred to PC via IEEE 1394a and saved in uncompressed TIF format in 2592 (H) \(\times\) 1944 (V) resolution in 12 bits.

Figure \ref{fig:fig1}(b) illustrates the typical information obtained at the exit surface of the MCMMF in the form of wavelength dependent speckle patterns obtained from the individual fiber cores. Each pattern corresponds to a superposition of higher order fiber modes that is dependent on the wavelength, the length of the fiber and the angle of incidence. All fiber cores are slightly different and local variations in the material properties, strain, impurities and other random structural elements give rise to an individual set of speckle patterns for each  fiber core in the array. The speckle patterns for every wavelength are stored into a multispectral transmission matrix for every core, which in principle allows retrieval of spectral information from arbitrary superposition states using a number of different techniques. 
Spectra consisting of more than one wavelength component, as well as continuous spectra, result in a superposition of many speckle patterns. Analytical inversion techniques like Moore-Penrose pseudo-inversion can be employed to reconstruct the spectra from these superpositions, but their performance is strongly dependent on noise and appropriate regularization is needed. In this work we compare our DL approach with analytical inversion using Tikhonov regularization (TR) \cite{liutkus_imaging_2014}. Moreover, analytical inversion is limited to the oversampled regime, and its breakdown is observed at the Shannon-Nyquist sampling limit \cite{french_speckle-based_2017}. Compressive sensing (CS) extends reconstruction into the undersampling regime under conditions of sparsity, in our work CS was implemented for comparison to DL using the python package ``cvxpy'' \cite{cvxpy_rewriting}.

\begin{figure}[th]
	\centering\includegraphics[width=10cm]{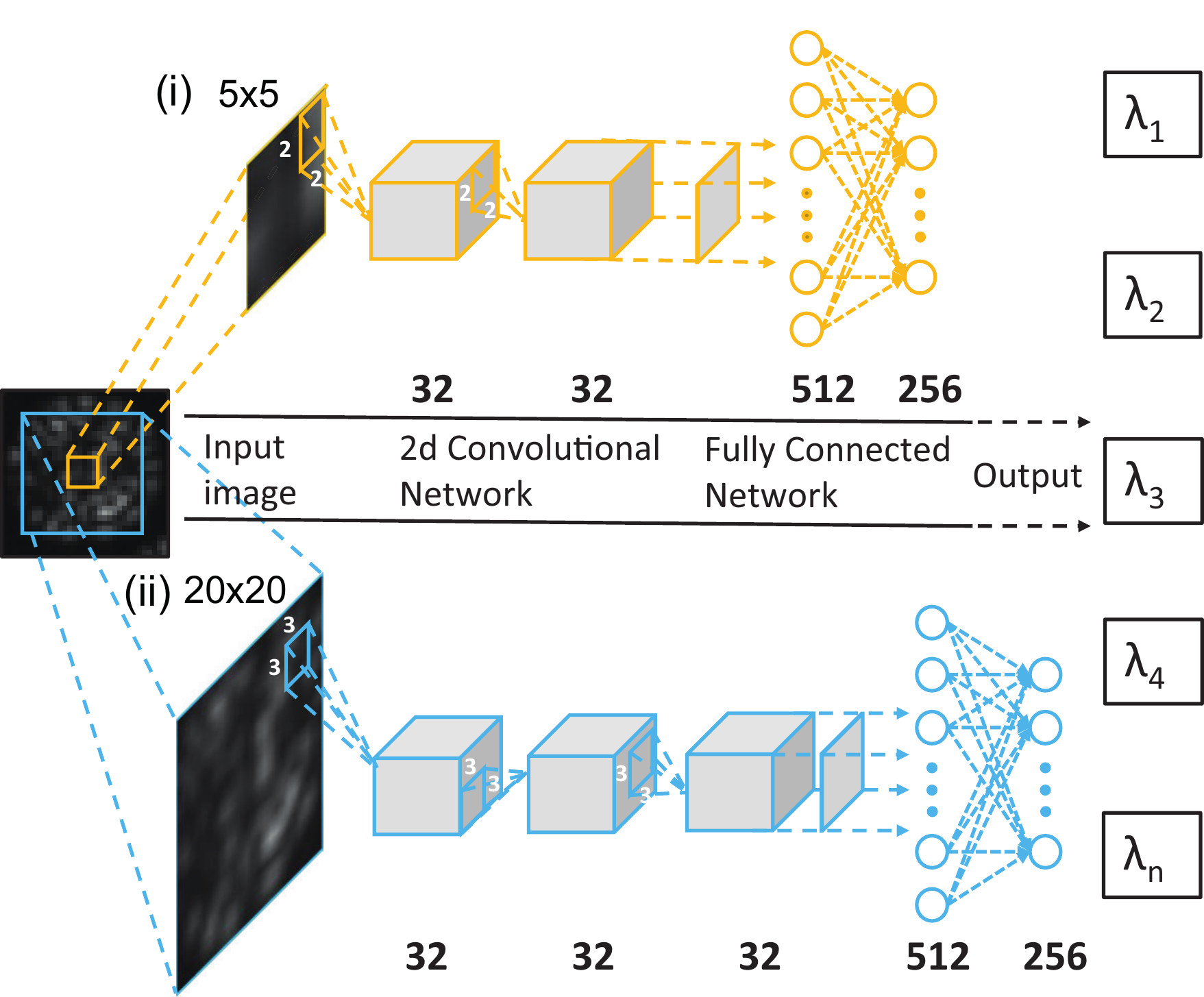}
	\caption{Neural network structures used in this study for pixel areas of (i) 5x5 pixels (Y/X = 0.58) and (ii) 20x20 pixels (Y/X =  9.30).}
	\label{fig:fig2}
\end{figure}

Spectral reconstruction via DL was implemented using a convolutional neural network (CNN), composed of a series of convolutional layers followed by two fully connected layers of 512 and 256 nodes with dropout regularization using 70\% keep probability. The final, dense output layer of 43 neurons represents the spectrum, where each neuron corresponds to a discrete wavelength channel. The size of the CNN was manually optimized for each tested sampling condition. For a region of interest (ROI) size of 5x5 pixels, the best performing network consists of two convolutional layers with each a 2x2 kernel (CNN (i) in Fig. \ref{fig:fig2}, yellow). On 20x20 pixels, a three-layer CNN with kernel size 3x3 throughout the network was found to perform best (CNN (ii) in Fig. \ref{fig:fig2}, blue). Each convolution is followed by batch normalization and a leaky ReLU activation layer, all layers use valid padding. We found that any type of pooling consistently reduced the reconstruction quality, so no pooling was performed. The networks were implemented in python using keras as frontend for tensorflow \cite{tensorflow2015-whitepaper}. 

To test the performance of the different approaches, multiple patterns were digitally added up together to simulate a real signal made of a given number, \(N_{\lambda}\), of nonzero wavelength components with randomly varying intensities. The images of the speckle patterns were cropped to various ROI sizes to achieve different regimes of oversampling and undersampling, as given by the ratio of the total number of calibrated wavelengths, Y, to the total number of pixels of the ROI, X. For each multimode fiber, a total dataset of 31,000 images was generated, of which 29,000 were used to train the neural network, 1000 served for validation during training and the remaining 1000 were used for the final evaluation. A plot demonstrating training convergence is given in Fig. \ref{figSI:fig7} in the Appendix.

\section{Results and discussion}
\subsection{Deep learning reconstruction quality}

\begin{figure}[th]
	\centering\includegraphics[width=\textwidth]{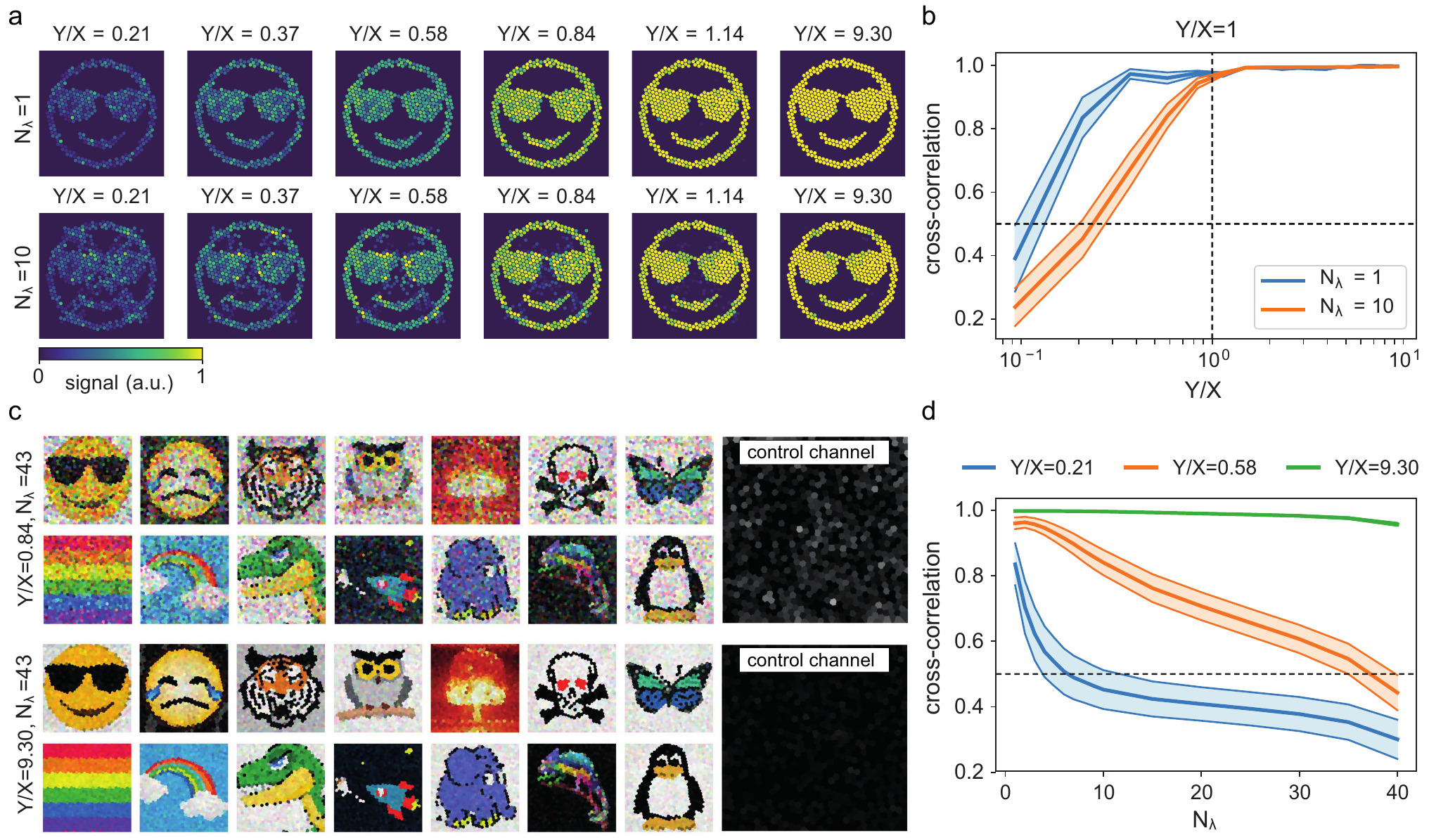}
	\caption{%
		a) Numerical illustration and b) calculation of reconstruction quality using DL for different sampling rates Y/X, for Nλ=1 and Nλ=10 non-zero wavelengths. One wavelength carries the encoded image (smiley), all other non-zero channels encode the image of a capital "X", which becomes slightly visible at low sampling rates due to cross-talk (see Supporting Information). 
		c) Numerical illustration of image reconstruction using DL for dense spectra (Nλ=42) showing 14 RGB images that are encoded in 42 wavelength channels, the 43rd, blank channel serves for cross-talk control. Reconstructions are shown for undersampling Y/X=0.84 and oversampling Y/X=9.30 regimes. 
		d) Cross-correlation with ground truth as a function of number of non-zero wavelengths in the spectrum for different sampling rates. Results in (b,d) are averaged over the whole fiber stack and for 100 spectra per fiber core. Light areas indicate the standard deviation of the data. Dashed line at Y/X=1 corresponds to Nyquist-Shannon sampling limit. Dashed line at a cross-correlation of 0.5 indicates the threshold below which the reconstruction is considered to have failed. Underlying full spectral data for (a) and (c) are presented in Fig. \ref{figSI:fig9} and Fig. \ref{figSI:fig10} of the Appendix. 
		All shown cliparts are from {www.openclipart.org} and public domain.}
	\label{fig:fig3}
\end{figure}

A direct numerical illustration of the reconstruction capabilities of the DL approach is shown in Fig. \ref{fig:fig3}. In Fig. \ref{fig:fig3}(a) and \ref{fig:fig3}(b) the performance is shown for different sampling regimes ranging from oversampling (\(Y/X=9.30\)) down to deep undersampling (\(Y/X=0.21\)). The cartoons in Fig. \ref{fig:fig3}(a) illustrate the quality of the reconstruction, where a smiley emoticon was used as the ground truth in a single wavelength channel. For a single non-zero wavelength \(N_{\lambda}\)=1, this is the only information contained in the spectrum. For the case \(N_{\lambda}=10\), nine other wavelengths in the spectrum are filled with a crossed-out symbol in the form of a capital ``X''. We see that in both cases, the reconstruction of the target is very good in the oversampling regime, while the quality becomes poorer below the sampling limit, \(Y/X<1\). For \(N_{\lambda}=10\) we can see the appearance of the cross in the image, indicating the presence of significant cross-talk between the spectral channels.

A quantitative analysis of this dependence of the reconstruction quality on the sampling rate is presented in Fig. \ref{fig:fig3}(b), where the cross correlation of the reconstructed spectrum with the input spectrum (ground truth) is plotted versus the sampling ratio Y/X. We can clearly see the main trends identified in the illustration, namely a good quality of reconstruction in the oversampling regime and a degradation in performance for \(Y/X<1\). In the undersampling regime, we see that the reconstruction improves for lower number of nonzero wavelengths, with reasonably good performance (defined as correlation \(> 0.5\)) for sampling rates as low as Y/X=0.21 for just a single non-zero wavelength. Clearly, DL is able to infer meaningful results under conditions in which the information density is sparse and where analytical inversion techniques show a complete breakdown \cite{french_speckle-based_2017}. In other words, DL is able to cross over far into the compressive sensing regime and therefore exhibits similar characteristics to a CS-based approach.
Having shown the strength of DL in the compressive sensing regime, it is of interest to investigate its capabilities in the regime of dense information, under conditions where the sparsity assumption underpinning CS ceases its validity. We start again with a numerical illustration in Fig. \ref{fig:fig3}(c) to visualize the amount of information that can be encoded through speckle images. Using Y=43 available wavelength channels, we encoded separately the red, green and blue (RGB) channels of 14 independent RGB images using the experimentally obtained transmission matrix. The remaining unused wavelength channel allows to assess the residual cross-talk. Raw RGB reconstruction data are given in the Supporting Information.

Figure \ref{fig:fig3}(c) shows the reconstructed RGB images obtained using our DL approach in either the regime of undersampling (Y/X=0.84) or oversampling (Y/X=9.30). In case of oversampling (\(Y/X>1\)), the neural network has much more input information to work with, which results in excellent image reconstruction quality and low residual cross-talk. For the undersampling regime, the images are still discernible but with significant reconstruction noise and cross-talk. These trends are again quantified in the accompanying analysis of Fig. \ref{fig:fig3}(d), showing the cross-correlation with the ground truth versus number of non-zero wavelengths. We see that the network output is almost perfectly correlated in the oversampled case (perfect reconstruction), which holds even for dense spectra, where signal is present in all wavelength channels (\(N_{\lambda}\)=43). As seen in Fig. \ref{fig:fig3}(c), for increasing number of wavelengths the effect of undersampling results in a rapid decrease of reconstruction fidelity. 
In the appendix we compare the same RGB data with reconstruction by TR and CS in the under- and oversampling regime and find that DL can compete with CS at a visual comparison and largely outperforms the analytical approach in the undersampled measurements.

\subsection{Comparison of different reconstruction techniques}

\begin{figure}[th]
	\centering\includegraphics[width=12cm]{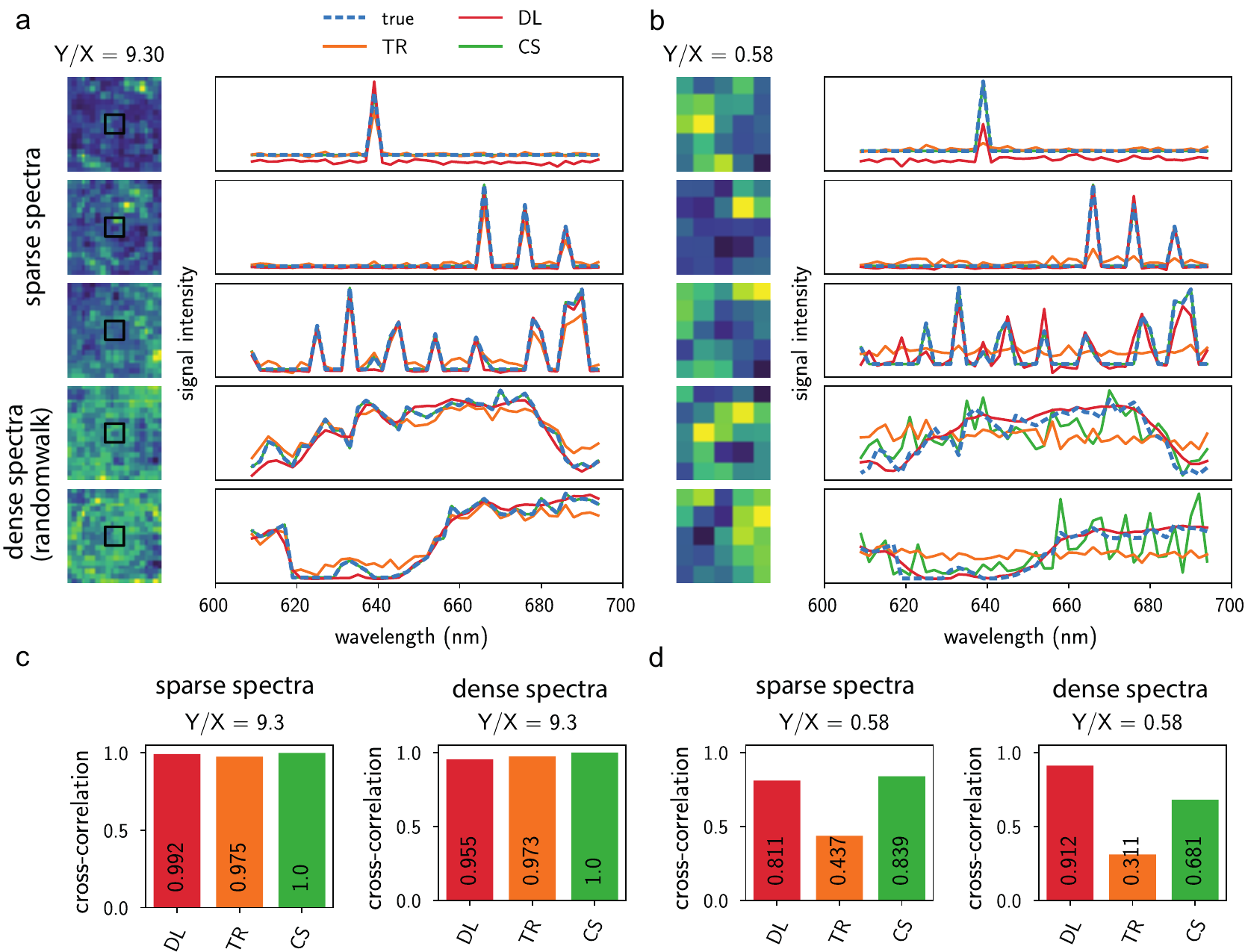}
	\caption{%
		a) Examples of speckle images and reconstructed spectral information for sparse (three top rows) and dense spectra (two bottom rows, generated by a random-walk like algorithm). Left column: oversampling regime with Y/X=9.3, right column: undersampling regime with Y/X=0.58.  The black box inside the speckle pattern shows the ROI used in the undersampling case. 
		b) Histograms comparing average cross correlations from 1000 randomly generated sparse (\(< 50\%\) sparsity) and dense (all wavelengths non-zero) spectra, obtained with deep learning (DL), Tikhonov regularization inversion (TR) and compressive sensing (CS).}
	\label{fig:fig4}
\end{figure}

In Fig. \ref{fig:fig4} deep learning (DL) is compared directly with both the TR and CS reconstruction methods.  In this benchmark, 1000 randomly generated spectra were generated numerically from the experimental transmission matrix. Figure \ref{fig:fig4}(a) shows the oversampling case (Y/X=9.30) corresponding to an ROI of 20x20 pixels, while Fig. \ref{fig:fig4}(b) shows the undersampling case (Y/X=0.58) corresponding to only 5x5 pixels. Several examples of typical spectra are shown (blue dash, ground truth), together with their corresponding reconstructions using DL (red), TR (orange) and CS (light blue). The lower two examples correspond to continuous spectra with a high density of spectral information.

Figure \ref{fig:fig4}(c) and \ref{fig:fig4}(d) gives the full quantitative analysis of the average of the cross-correlation between each of the 1000 randomly generated spectra (ground truth) and its respective reconstruction. In the oversampling regime, all methods perform well with correlation values \(>0.95\). In the undersampling regime TR fails completely (average cross-correlation \(< 0.5\)) as it can be seen to produce a mostly flat spectrum for all cases irrespective of the spectral shape. DL yields a very good performance and even clearly outperforms CS for dense spectra in the undersampling case. 

The slightly weaker performance of DL compared to CS in the oversampling case can be explained by the statistical training procedure in DL, while CS on the other hand is an analytical approach, yielding generally a close-to-optimum solution – however at a significantly higher computational cost compared to the neural network reconstruction, as discussed further below. In the undersampling case we observe that DL tends to result in inferred spectra that are smoother than the original input spectrum, whereas CS results in more spikes in the spectra even when the input is smooth. 

\begin{figure}[th]
	\centering\includegraphics[width=\textwidth]{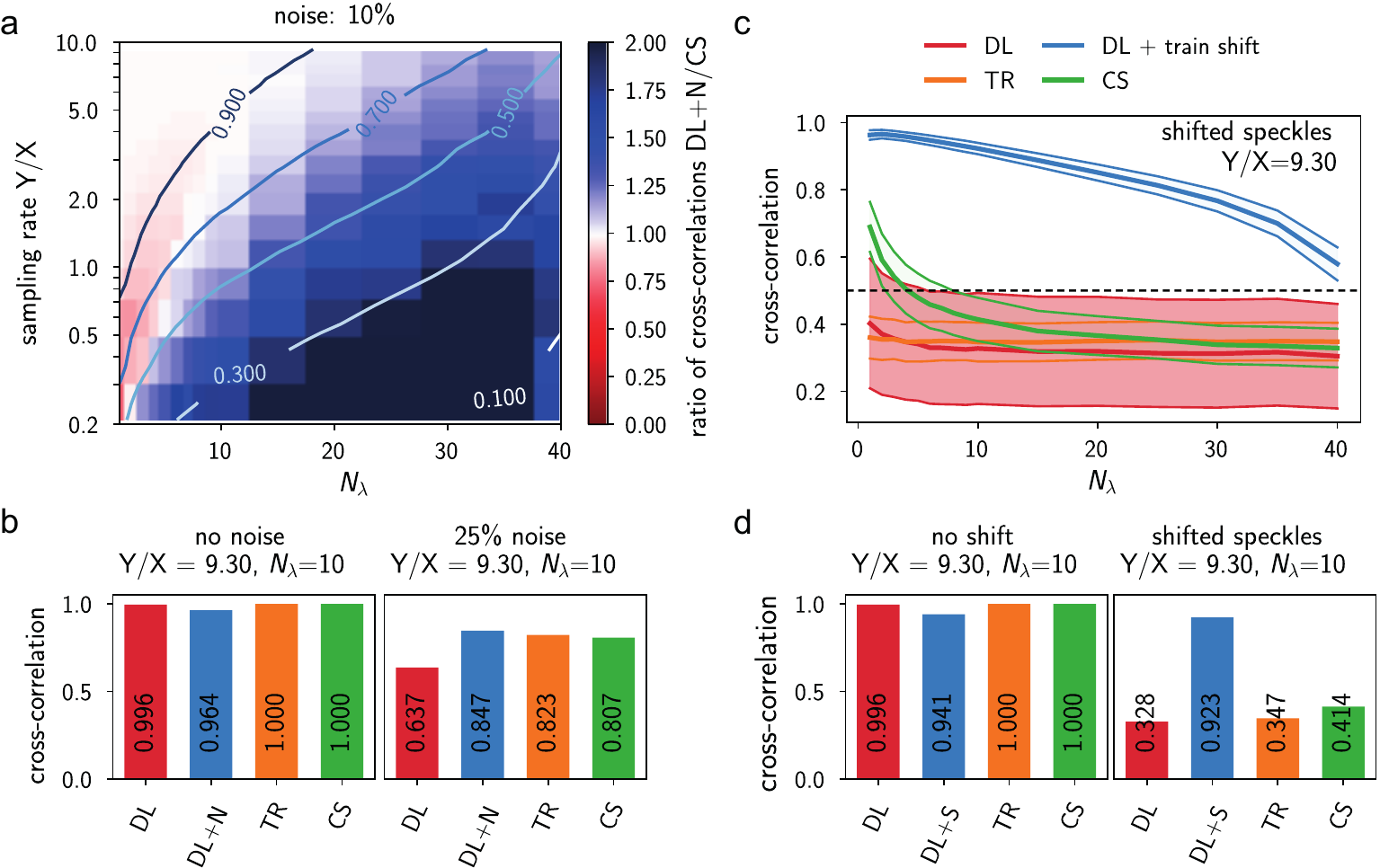}
	\caption{%
		a) Map showing ratio of reconstruction quality (cross-correlations) for deep learning trained on noisy data (DL+N) over compressive sensing (CS) for 10\% added noise. Blue indicates DL+N better than CS, red indicates CS better than DL+N. Contour lines indicate the cross-correlation of DL+N speckle reconstruction. 
		b) Calculated cross-correlations without noise and in presence of 25\% noise. DL+N outperforms CS over a large part of parameter space where the cross-correlation \(< 0.9\). 
		c) Robustness of the reconstruction against shifts of the speckle patterns by one pixel in a random direction. 
		d) Calculated cross-correlations without shift and with shift of 1 pixel. DL can be trained on data including shift (DL+S), which renders the method very robust in such scenario, largely outperforming TR and CS.}
	\label{fig:fig5}
\end{figure}

The previous tests considered speckle reconstruction using a multispectral transmission matrix in the complete absence of noise. In a real-life scenario, one can expect some level of noise to be present in the image, either shot noise, electronic camera noise or other non-specific backgrounds. An imaging system may also experience some drift caused by environmental effects, such as vibrations and thermal variations. To assess the robustness of the different approaches against typical perturbations, we compare in Fig. \ref{fig:fig5} the respective performance of DL, CS and TR. While CS and TR are based on analytic methods which are intrinsically inflexible to variations, DL has the advantage of allowing some level of adaptability through the choice of training data.

\subsection{Robustness against noise and spatial shifts}
To investigate the adaptability of the DL approach, we trained the neural network using noisy and spatially shifted training data with the aim of making it more robust against these effects. To account for the effect of noise, we added normally distributed random intensity noise to every pixel of the speckle pattern. 
Figure \ref{fig:fig5}(a) shows a parameter map showing the relative performance of DL trained on noisy data (DL+N) against CS, which is quantified as the ratio of their respective cross-correlations with the input spectra. Noise-adapted DL consistently outperforms CS over a large parameter-space. Blue regions indicate a superior DL performance. DL always performs better for relatively dense spectra with \(N_{\lambda}>15\). Even in most of the ``white'' parts of the colourplot, DL is outperforming TR and CS by at least some percent (see bar plots in Fig. \ref{fig:fig5}(b)). CS is better performing only on sparse data at very low sampling rates. Furthermore TR contains a free parameter which has to be empirically adjusted to match a given noise level in order to achieve the specified performance \cite{liutkus_imaging_2014}. The neural network trained on noisy data (DL+N) outperforms the normal DL when dealing with noisy spectra. This increased adaptability of the DL+N neural network comes at the cost of a relatively small hit on performance when dealing with noiseless data as seen in Fig. \ref{fig:fig5}(b).

In principle, the performance of CS and TR on noisy data can be improved by adding an additional denoising step prior the spectral reconstruction, which of course adds additional computational cost. Since we are interested in developing a real-time-ready approach, computationally expensive image treatment on each speckle should be avoided. Thus, our analysis evaluates, how the different approaches can handle noisy data ``out of the box''.

A larger drift of the CCD or of the fiber would completely change the transmission matrix of the system and hence would require a full re-characterization of the setup. In the following we test whether deep learning can adapt to such situations, where analytical methods and compressive sensing break down definitely.
To simulate spatial drift, the cropped ROI was randomly shifted by one pixel in arbitrary directions. We compare in Fig. \ref{fig:fig5}(c) and \ref{fig:fig5}(d) the performance of networks that were trained with and without spatial shift. If the neural network is trained without such perturbed data, the performance of DL decreases, just like for the conventional methods TR and CS. However, if the training datasets contain accordingly perturbed data, DL shows a distinct performance boost at the reconstruction from non-perfect speckle images. In the case of shift (Fig. \ref{fig:fig5}(c) and \ref{fig:fig5}(d)), accordingly trained DL dramatically outperforms the other techniques on perturbed data, while again the loss of performance on unshifted data is relatively small.

We note that the training set might not only contain data accounting for shifts of the CCD, but also for shifts of the fiber or variations in the incident angle of light. All these effects would affect the speckle patterns in a deterministic way and could be corrected by an accordingly trained artificial neural network, whilst this is impossible with an analytical method.

\subsection{Real time reconstruction of hyperspectral images}

\begin{figure}[th]
	\centering\includegraphics[width=\textwidth]{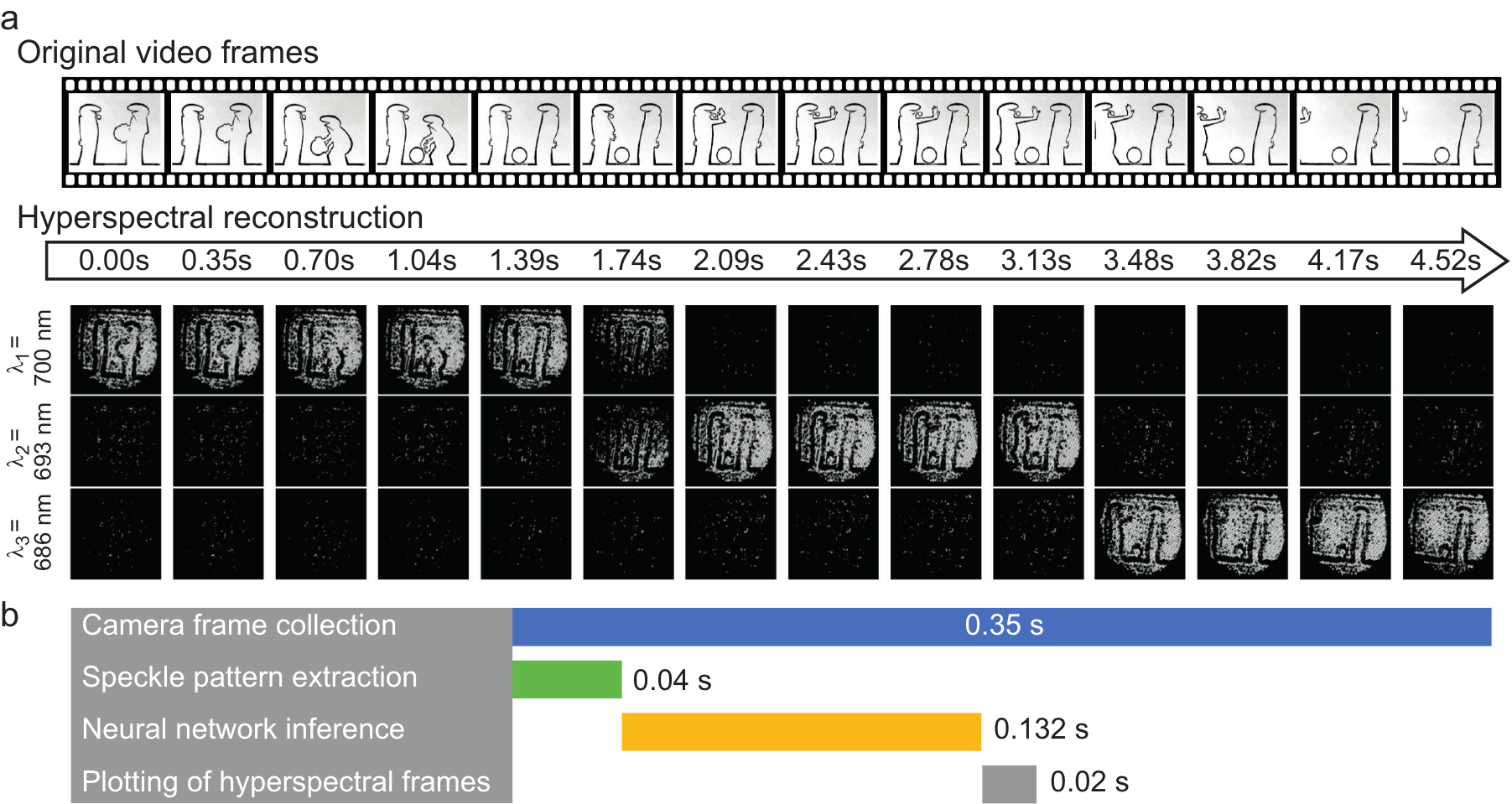}
	\caption{%
		a) Real-time speckle-based hyperspectral video reconstruction via DL. A video is projected on the fiber bundle using an SLM in amplitude-modulation configuration. During playback the wavelength of the projecting light is changed. Top: Original frames of the input video (see also Visualizations). Bottom: Spectral reconstruction of three wavelength channels of the full multi-core fiber (approx. 2700 fiber cores). 
		b) Bar graph showing the timings of the different execution steps. Full visualizations are included in the Supporting Materials of this study. La Linea, with permission, copyright CAVA/QUIPOS.}
	\label{fig:fig6}
\end{figure}

Finally, apart from its universal applicability and superior stability, the most important achievement of using DL over CS is its reconstruction speed. We therefore demonstrate that our deep learning based hyperspectral reconstruction method is capable of real-time image processing. To this end we projected a grayscale video via an SLM onto the MCMMF. During video playback we randomly changed the wavelength of the illumination using the AOTF.
We analyzed in real time the 2700 speckle patterns for each core, captured by the CMOS using artificial neural networks (one network for each fiber), which were trained in advance and pre-loaded into RAM. Figure \ref{fig:fig6}(a) shows frames from the original video data in the top row, and the three first wavelengths of the spectral reconstruction in the bottom rows (for full videos see Visualizations). Upon switching of the wavelength, the reconstructed image changes channel with limited cross-talk between channels. At 1.74s, the AOTF was switched during the frame acquisition, briefly resulting in two spectral components. We note that the reconstruction quality is a bit worse compared to the synthetic data shown before, which is due to non-perfect intensity stability of the setup and some spurious residual wavelength correlations in the used, shorter (2.54cm) fiber. Further visualizations with the ``La Linea'' video and with an ``Eclipse'' video are also available.
Finally, using microsecond-fast frequency sweeps with the AOTF, we were able to transmit spectrally broadened wavelength windows through the fiber bundle. Neural network based, multi-wavelength hyperspectral image reconstruction of experimental real-time data is demonstrated in some of the Visualizations in the supplementary material.

The training of the networks for 2700 fiber cores takes about 8 hours on an Nvidia Quadro P6000 GPU. This is a one-time procedure since the speckle patterns can be maintained stable over long times. The hyperspectral image reconstruction itself is the time-critical process in many applications. The trained neural networks are capable to reconstruct all 2700 fiber cores in only 132 ms on an Intel i7-3770 CPU. In comparison, CS requires around 2 minutes per frame for processing of the full fiber bundle. TR should potentially be even faster than deep learning, but interestingly, we found that in our python implementation, TR is around 1.5 times slower than the neural network inference (190\,ms per video frame), which we attribute to the relatively slow python code, whilst the neural networks are evaluated through the highly optimized tensorflow backend. Hence it should be possible to further accelerate TR.
Visually, all three methods perform similarly on the experimental video reconstruction.

Figure \ref{fig:fig6}(b) shows the durations of the steps of DL hyperspectral image reconstruction on the Intel i7-3770 CPU, working with 32GB of RAM for network pre-loading. Including the preprocessing and rendering of the final hyperspectral images, the algorithm in its current state can reconstruct about 5 full images per second, which is faster than the total acquisition and transfer time of our 12bit CMOS camera (about 0.35s). In supplementary Visualizations, we also show \(>\)5fps reconstruction rate, using faster 8bit CMOS readout, which becomes essentially limited by the neural network inference speed.

There is significant scope for further increasing the reconstruction speed of DL by performance, in a first step  already by optimization of the code. While not all 2700 networks fitted in the memory of our GPU, a benchmark on 1000 preloaded networks showed that the GPU offers around \(\times\)2.5 speed improvement, multi-GPU platforms will be accordingly faster.
Training networks that each reconstruct speckles for multiple fibers in parallel can potentially result in a further significant performance boost and reduce the memory requirement of the approach, as is shown in figure~\ref{fig:appendix_multi_fiber_net} of the appendix.
Even higher speed can potentially be obtained by developing hardware implementations of the networks, for example based on Field Programmable Gate Arrays (FPGAs) \cite{accelerating-deep-convolutional-neural-networks-using-specialized-hardware}. 
With respect to our current implementation using Python, a more direct hardware/software communication could also straightforwardly increase the frame collection rates, which however is outside the scope of this proof of principle study.

\section{Conclusion}

In conclusion, we have shown that with deep learning, a multicore multimode fiber bundle can be used as a real-time hyperspectral camera, robust to noise and spatial shifts. Using wavelength- and fiber-core dependent speckle patterns, we have used deep learning to cope with large amounts of data at a video rate of several frames per second on conventional computer hardware. The imaging spectrometer and deep learning technique are versatile by design, and the calibrated wavelength range can be tailored to specific applications. The approach can be easily scaled to any number of wavelength channels, to desired spectral resolution via the length of the multicore fiber bundle and with respect to imaging spatial resolution. Deep learning in combination with speckle spectrometry enables a new class of low-cost, compact hyperspectral imaging systems with real-time data processing capabilities.

\section*{Supplementary materials}

Visualizations 1-4:  ‘La Linea’ short animation demonstrating the spectral deconstruction using the MCMMF and deep learning algorithm for several AOTF wavelength sweep sequences.

Visualization 5:  ‘Eclipse’ short animation demonstrating the spectral deconstruction using the MCMMF and deep learning algorithm.

\section*{Funding}
German Research Foundation (DFG) Research Fellowship (WI 5261/1-1); EPSRC (EP/J016918/1).

\section*{Acknowledgments}
We gratefully acknowledge the support of NVIDIA Corporation with the donation of the Quadro P6000 GPU used for this research. La Linea © CAVA/QUIPOS. La Linea usage rights granted to the University of Southampton for the purpose of this research has been approved by Quipos, Osvaldo Cavandoli's worldwide licensor.
introduction to DBSCAN, and to Laurent Daudet for help with compressive sensing. All data supporting this study are openly available from the University of Southampton repository (DOI: 10.5258/SOTON/D0942).

\section*{Appendix}

\begin{figure}[h]
    \centering\includegraphics[width=8cm]{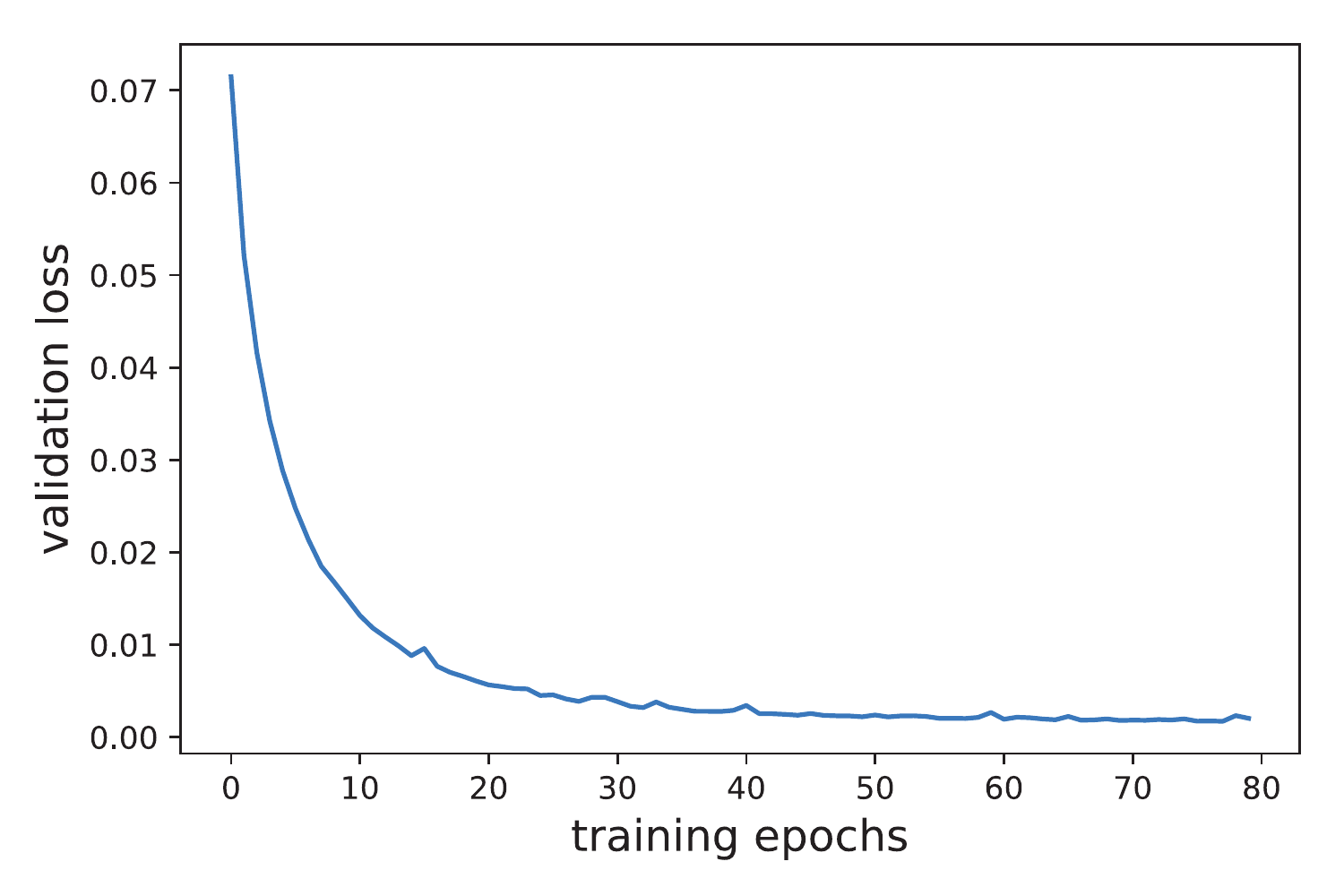}
    \caption{%
        Least squares loss on validation data (1000 spectra) as function of training epoch (during one epoch, the full training dataset is used once for optimization of the network weights and biases). Training is done on a set of 29000 spectra.}
    \label{figSI:fig7}
\end{figure}

\begin{figure}[h]
    \centering\includegraphics[width=11cm]{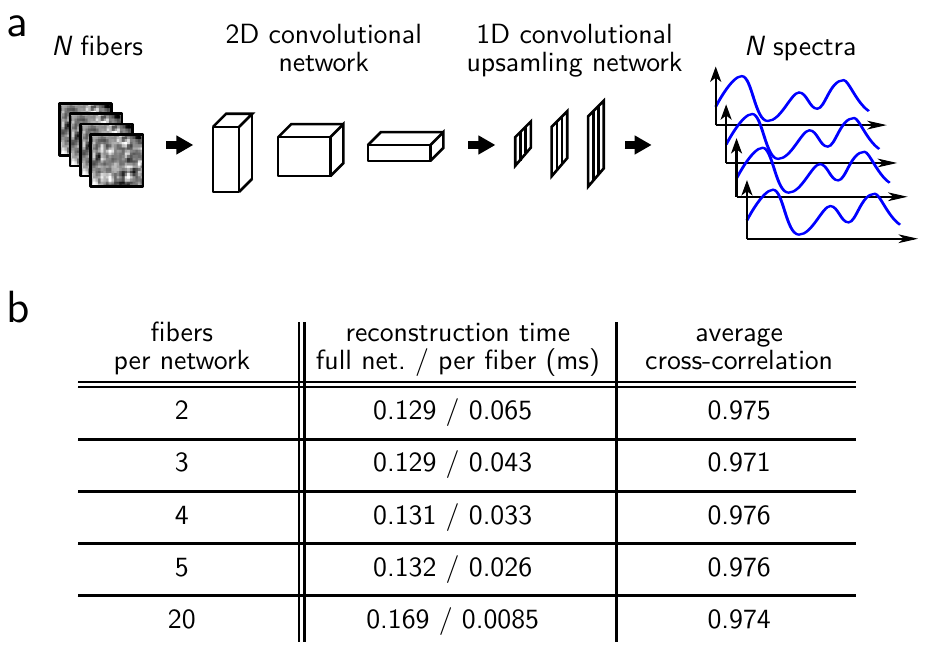}
    \caption{%
        Training multi-channel 2D convolutional -- 1D convolutional upsampling networks on multiple fibers per network.
        a) architecture of a multi-fiber network: speckle images from \(N\) different fibers are fed into the network and translated into \(N\) spectra at the output. 
        b) timing and reconstruction quality for a fixed network architecture for increasing number \(N\) of reconstructed fibers (corresponding to the number of input and output channels). Trained on 10,000\(\times N\) speckles. On 32GB RAM, training data for up to 20 fibers could be kept in memory. The trained networks show no significant difference in reconstruction quality, while the timing per fiber reduces almost linearly. The Memory requirement of each pretrained network is approximately constant, hence required RAM linearly decreases with the number of reconstructed fibers per network. 
        }\label{fig:appendix_multi_fiber_net}
\end{figure}

\begin{figure}[h]
    \centering\includegraphics[width=11cm]{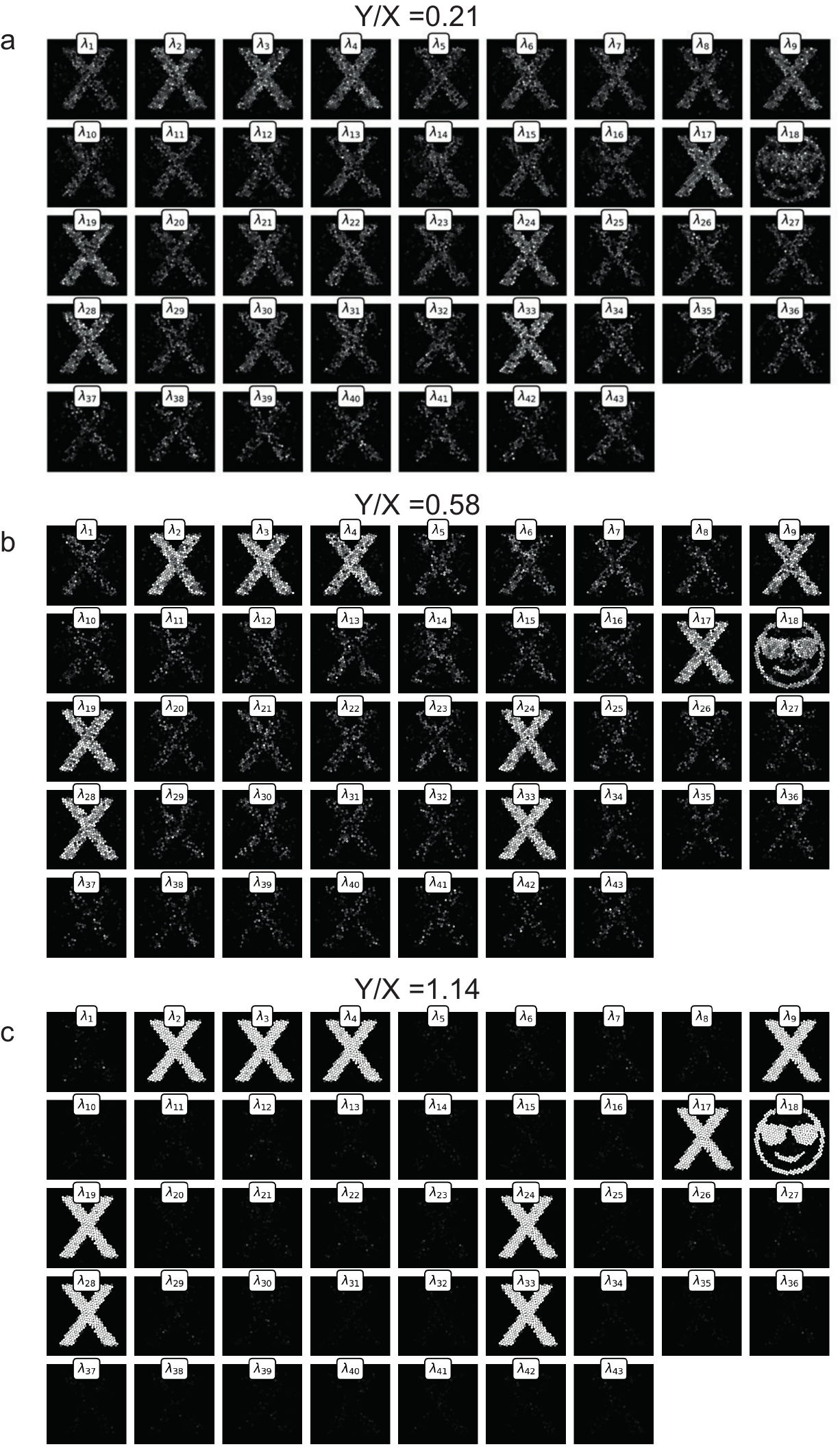}
    \caption{%
        Reconstruction of all spectral channels with 10 channels containing images at a sampling rate of (a) Y/X=0.21, (b) Y/X=0.58, (c) and Y/X=1.14. Channels carrying information are 2; 3; 4; 9; 17; 18; 19; 24; 28; 33.}
    \label{figSI:fig9}
\end{figure}

\begin{figure}[h]
    \centering\includegraphics[width=\textwidth]{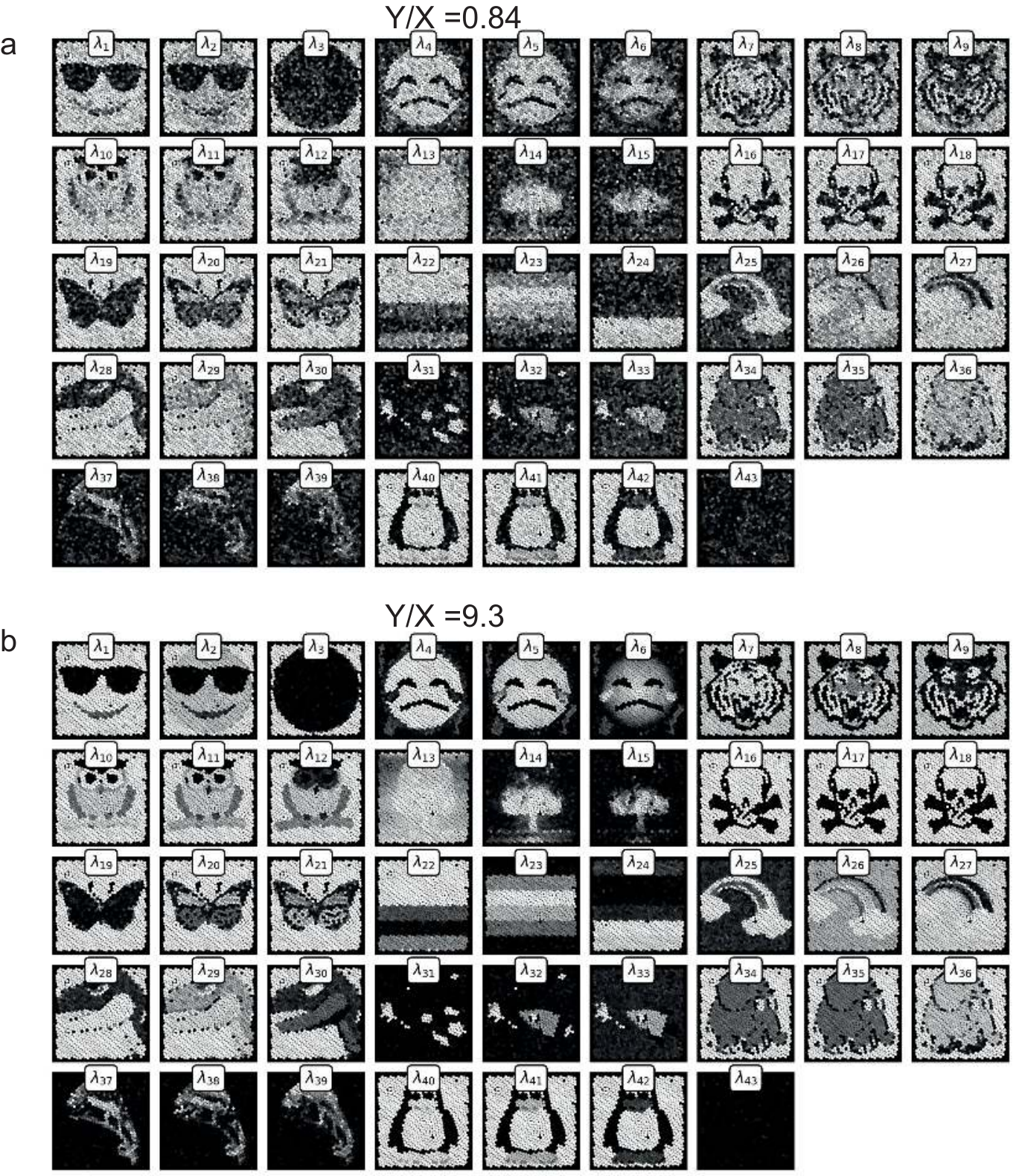}
    \caption{%
        a) Encoding and reconstruction of 14 RGB images in the speckle patterns of the multi-core fiber. Raw reconstructions of the spectral channels, containing the red, green and blue parts of the color images, at a sampling rates of (a) Y/X=0.84 and (b) Y/X=9.3.}
    \label{figSI:fig10}
\end{figure}

\begin{figure}[h]
	\centering\includegraphics[width=10cm]{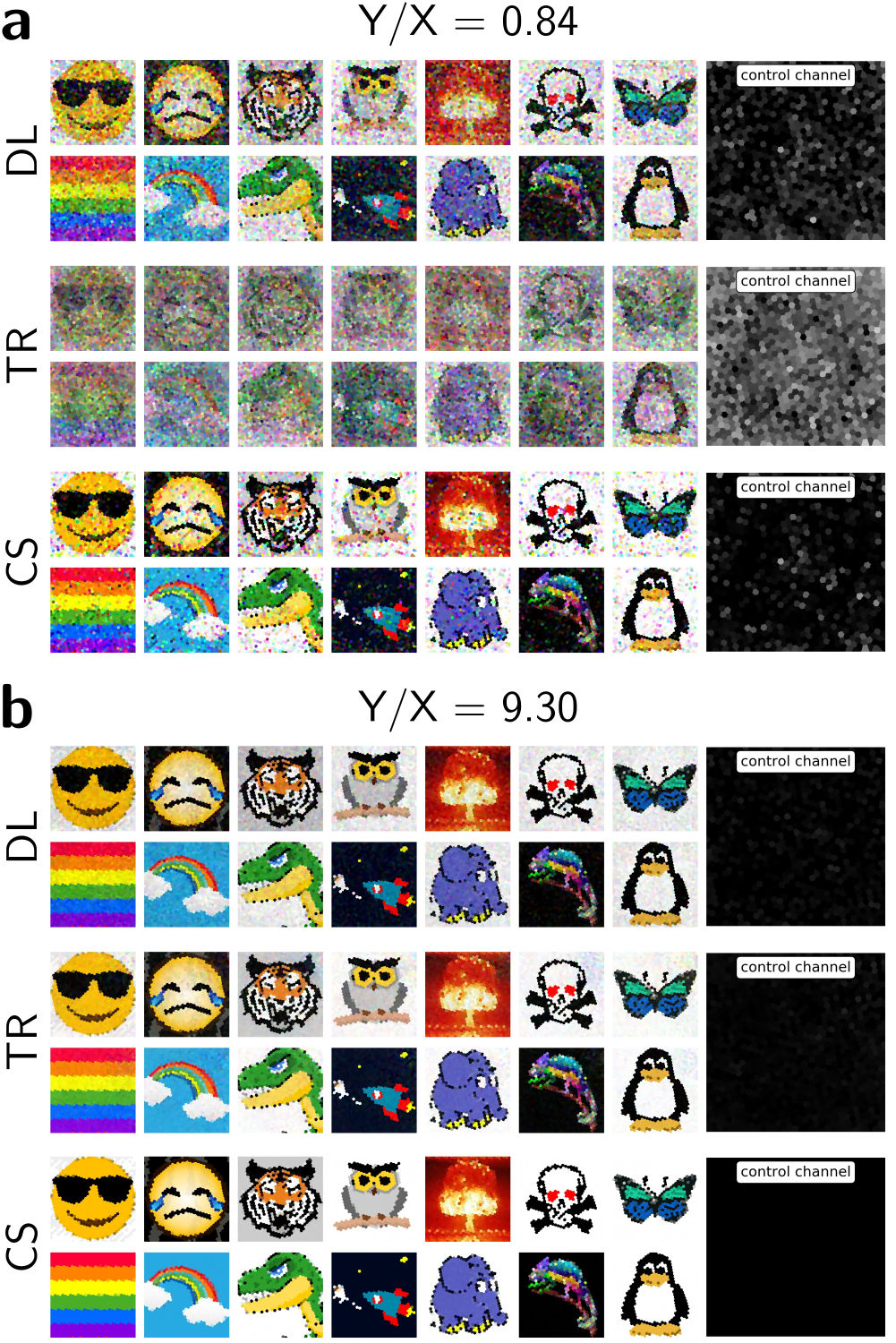}
	\caption{%
		Visual comparison of reconstruction quality of the different methods (deep learning [DL], Tikhonov regularization [TR] and compressive sensing [CS]) on the RGB dataset (see also figure~3). a) sampling rate of Y/X=0.84 and (b) Y/X=9.3.}
    \label{figSI:fig11}
\end{figure}

\end{document}